\documentclass[11pt]{iopart}
\expandafter\let\csname equation*\endcsname=\relax
\expandafter\let\csname endequation*\endcsname=\relax

\usepackage{iopams}
\usepackage{amsmath,amsfonts,amssymb,latexsym,graphicx,color,epstopdf}
\usepackage[utf8]{inputenc}
\usepackage{ulem}
\usepackage[margin=10pt,font=small,labelfont=bf,labelsep=endash]{caption}
\usepackage{stackrel}
\usepackage{pifont}

\definecolor{lightblue}{rgb}{0.176471,0.396078,0.709804}
\newcommand{\pd}[2][]{\frac{\partial#1}{\partial#2}}

\newcommand{\dd}[2][]{\frac{\mathrm{d}#1}{\mathrm{d}#2}}
\newcommand{\mc}[1]{\mathcal{#1}}
\newcommand{\mb}[1]{\mathbf{#1}}

\newcommand{\dx}[1]{\,\mathrm{d}#1}
\newcommand{\ov}[1]{\overline{#1}}
\newcommand{\nbxy}{\nabla_{\parallel}}
\newcommand{\rx}[1]{{\color{red}#1}}
\newcommand{\bx}[1]{{\color{lightblue}#1}}
\newcommand{\mf}[1]{\mathfrak{#1}}

\begin{document}

\title{Self-Similar Cusp Formation in Thin Liquid Films By Runaway Thermocapillary Forces}
\author{Chengzhe Zhou \dag\ and Sandra M. Troian\ddag \footnote[3]{Corresponding author (stroian@caltech.edu)}}

\address{\dag\ California Institute of Technology, 1200 E. California Blvd, Physics, MC 103-33, Pasadena, CA 91125, USA}

\address{\ddag\ California Institute of Technology, 1200 E. California Blvd., Applied Physics, MC 128-95, Pasadena, CA 91125 USA}

\begin{abstract}
Many physical systems give rise to dynamical behavior leading to cuspidal shapes which represent a singularity of the governing equation. The cusp tip often exhibits  self-similarity as well, indicative of scaling symmetry invariant in time up to a change of scale. Cusp formation can even occur in liquid systems when the driving force for fluid elongation is sufficiently strong to overcome leveling by capillarity. In almost all cases reported in the literature, however, the moving interface is assumed to be \textit{shear-free} and the operable forces orient exclusively in the direction normal to the advancing boundary. Here we focus on a system in which a slender liquid film is exposed to large thermocapillary stresses, a system previously shown to undergo a linear instability resembling microlens arrays. We demonstrate by analytic and numerical means how in the nonlinear regime these surface forces undergo self-similar runaway behavior leading to cusp formation with a conical tip whose slope can be prescribed from the analytic relation derived. On a fundamental level, this finding broadens our understanding of known categories of flows capable of cusp formation. More practically, the system geometry proposed offers a potentially novel lithographic method for one-step non-contact fabrication of cuspidal microarrays.
\end{abstract}


\submitto{New Journal of Physics (26 July 2018)}
\maketitle

\section{Cusp formation in physical systems}
\begin{figure}
\centering
\includegraphics[scale=1.0]{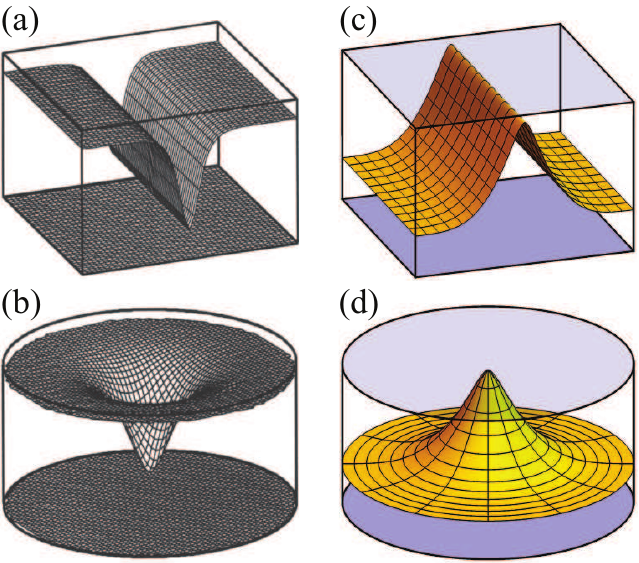}
\caption{Liquid (a) line type and (b) point type cusp formation in a thin film subject to a negative disjoining pressure from van der Waals forces that promotes dewetting of the film from the bottom solid substrate \cite{WB00}. Liquid (c) line type and (d) point type conical cusp formation caused by thermocapillary forces which draw fluid away from the lower warm substrate toward the colder top substrate, described in more detail in the text.}
\label{fig:touch_concept}
\end{figure}

Despite that capillary forces always act to repress regions of high curvature, nature nonetheless finds clever ways of forming and sustaining cusps in many physical systems. In fact, cusps are rather ubiquitous and occur in such diverse phenomena as thermal grooving at grain boundaries \cite{M57}, surface diffusion and pinchoff in annealed or sintered systems \cite{BBW98}, complex plasma formations \cite{SM09}, wavefront propagation in systems described by the linear \cite{YB12} or nonlinear Schr$\ddot{o}$dinger equation \cite{AA11}, critically charged droplets \cite{BT11}, microbranching instabilities in fast moving cracks \cite{KF15}, line attractor states in neural computation models \cite{XT17} and many more. A recent delightful book by J. Eggers \cite{EF15} describes as well the complex dynamics governing cusp formation in many liquid systems including thread and droplet breakup, Hele-Shaw sink flow, and thin film rupture caused by a negative disjoining pressure which induces a dewetting process \cite{BBW98,ZL99,TABD15}. The latter system is sketched in Fig. \ref{fig:touch_concept} (a) and (b) where the receding air/liquid interface is observed to form a cuspidal curve.

In these and other systems \cite{E93,E01,Z04,BZ09,KS17}, the apical region of the evolving cusp exhibits self-similar behavior characterized by universal exponents, some of which have been confirmed experimentally \cite{CE06,CN02,PL09,MS14,VA16}. The resulting power laws stem from scaling symmetries that are invariant in time up to a change of scale. In almost all cases reported in the literature, however, the moving interface is assumed to be \textit{shear-free} where the operable forces orient exclusively in the direction normal to the advancing boundary. The interface therefore experiences no shear forces and plays no active role in corralling fluid into a sharpened tip. And while there have been observations of cusp formation leading to tip streaming in droplet systems subject to interfacial shear from surfactant concentration gradients \cite{K12,K15}, the dynamics of cusp formation there remains an unsolved problem.

To explore cuspidal formation driven by shear forces at a free interface, we here focus on a thin film system designed to elicit self-reinforcing thermocapillary stresses at the air/liquid interface. We analyze the dynamics by which the ensuant self-similar process gives rise to fluid elongations shaped like liquid cusps whose conical tips further promote self-focusing. Shown in Fig. \ref{fig:touch_concept} is an example of a thermocapillary (c) line and (d) point cusp caused by runaway thermocapillary forces. While Figs. \ref{fig:touch_concept} (a) and (b) depict cusp formation arising from forces exclusively oriented normal to the free interface (disjoining pressure counterbalanced by capillary pressure), Figs. \ref{fig:touch_concept} (c) and (d) depict formation of a cusp from thermocapillary (shear) forces which orient parallel to the moving interface. An additional distinction between these two thin film systems is that the thermocapillary problem exhibits multiscale dynamics in the apical region, which although quite interesting, considerably complicates any stability analysis.

\begin{figure}
\centering
\includegraphics[scale=1.0]{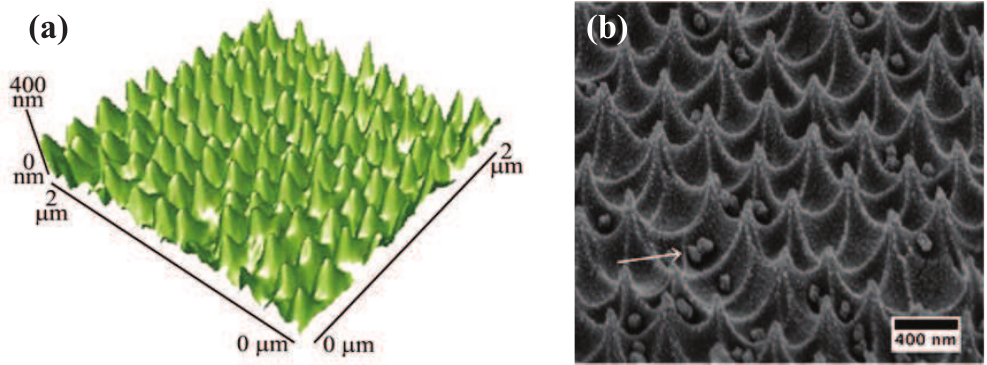}
\caption{Cuspidal arrays. (a) SEM micrograph of plasma etched substrate for super antireflective coatings \cite{WA16}. (b) AFM image of cicada wing \cite{WC13}.}
\label{fig:biomimetic}
\end{figure}

Aside from such fundamental considerations, there is practical motivation for this study as well. We are interested in exploring thermocapillary based techniques for patterning thin liquid films which can be rapidly solidified in situ. The system geometry examined in this work offers a potentially novel lithographic method for one-step non-contact fabrication of cuspidal microarrays. This development can facilitate design and manufacture of specialty microarrays such as biomimetic cuspidal substrates. Two recent important examples include infrared (IR) antireflective moth eye surfaces patterned with quintic cusps for eliminating Fresnel reflections in the mid-IR \cite{S91,WA16}, as shown in Fig. \ref{fig:biomimetic} (a), and superhydrophobic, self-cleaning antimicrobial surfaces mimicking the surface of a cicada wing \cite{WC13,NL17}, as shown in Fig. \ref{fig:biomimetic}(b). Such surfaces can likely be architected using thermocapillary forces to create substrates for which form follows function i.e. imprinted cuspidal shapes relate directly to their intended function.

Our group has previously demonstrated experimentally \cite{MT11} how patterned thermocapillary forces can be used to sculpt nanofilms into liquid microlens arrays, which are then solidified rapidly in situ. The resulting ultrasmooth surfaces are ideally suited to micro-optical applications such as beam shaping. The analysis presented in this work now suggests that were the microlens configuration allowed to evolve further in time, the system would transition to a microcuspidal array. The local analysis presented in this work indicates how initial protrusions of any sort, whether triggered by the linear instability \cite{DT09,DT10} or triggered by large amplitude perturbations  \cite{DT09Conf,MT11}, are expected to evolve into individual or array-like cuspidal patterns.

The outline of this work is as follows. In Section \ref{tcmodel} we present the thin film evolution equation for a molten Newtonian which are imposed by thermal conduction across a very slender confined system. This system gives rise to an initial linear instability whose wavelength characterizing the fastest growing mode is subsequently used to rescale the original equation to dimensionless form. Further rescaling to parameter-free form yields an equation belonging to the general class of so-called gradient flows, which in this case also contains a \textit{virtual} singularity where thermocapillary stresses diverge to infinity. In Section \ref{gradflowstability}, it is shown that this evolution equation equation does not support any stable stationary states since for small excursions about a stationary state there exist states of even lower energy. This demonstration proves that the dynamics incurred by the confined geometry imposed on the film leads to a runaway thermocapillary process in which the liquid can reduce its free energy by advancing ever closer to the top colder substrate. In Section \ref{numsoln}, 2D and 3D numerical solutions of the nonlinear evolution equation reveal stable formation of a cusp with a conical tip that undergoes continuous sharpening by a self-similar process exhibiting power law growth in the tip speed and tip curvature. In Section \ref{asymptotics} we present an asymptotic analysis of the apical region about the virtual singular point which reveals the presence of a stable fundamental mode which appears to act an attractor state. Various measures characterizing this fundamental mode are shown to be in excellent quantitative agreement with numerical simulations of the late time asymptotic behavior of the apical region. The asymptotic analysis also reveals  a simple relation for the conical tip slope that can be used to prescribe the shape for experimental applications. In Section \ref{conclusion}, we conclude with some final thoughts on how these findings may held advance a novel lithographic method for fabrication of specialty cuspidal microarrays.

\section{Long wavelength thermocapillary model for growth of protrusions by increasing interfacial shear forces}
\label{tcmodel}
\begin{figure}
\centering
\includegraphics[scale = 0.8]{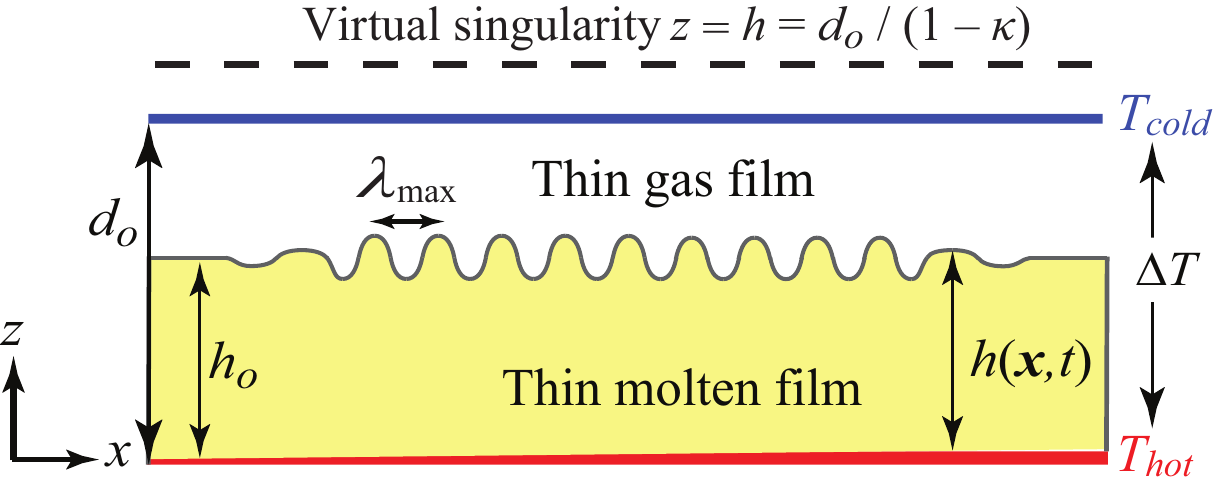}
\caption{Sketch of a linearly unstable thin molten film overlay by a gas layer. A large vertical temperature gradient is enforced by application of a uniform temperature difference $\Delta T = T_{\mathrm{hot}} - T_{\mathrm{cold}}$ maintained across a small gap width $d_o$, typically measuring about a micron or less. Estimated temperature gradients $\Delta T/d_o$ extracted from experiments reported in the literature \cite{DT10} range from about $10^6 - 10^8$ $^\textrm{o}\textrm{C/cm}$. As described in the text, the governing equation for the thin film contains a virtual singularity at $h=d_o/(1 - \kappa)$ designated by the dashed line, which lies beyond the top cold substrate since the ratio $\kappa$ is always less than one.}
\label{fig:TCexp}
\end{figure}

A theoretical model has previously been derived \cite{DT09,DT10} to describe the evolution and stability of a gas/liquid interface for the system sketched in Fig. \ref{fig:TCexp}.
A molten nanofilm of initial uniform or average thickness $h_o$ overlay by a slender gas film is confined within a very narrow gap $d_o$ (typically less than a micron) by two opposing substrates maintained at a uniform temperature difference $\Delta T = T_{\mathrm{hot}} - T_{\mathrm{cold}} > 0$. The model assumes that the film thickness is much smaller than any characteristic lateral scale, that inertial forces are negligible, and that the viscosity of the film $\mu = \mu (T_{\mathrm{hot}})$ is relatively constant given the very small gap width dimension $d_o$. The film thickness is restricted to the range $0 < h(\mb{x},t) < d_o$ where $\mb{x}=(x,y)$. Since for single component fluids the variation in surface tension $\gamma$ with temperature $T$ given by $d\gamma/dT$ is a negative quantity, those portions of the liquid film which are closer to the cold substrate experience a colder temperature and consequently a higher local value of surface tension. Temperature variations along the liquid interface therefore give rise to spontaneous interfacial thermocapillary stresses given by $\nabla_{\|} \gamma = (d \gamma/dT) \nabla_{\|} T$, which act to pull liquid from warmer to cooler regions of the film. Within the long wavelength approximation, the operator $\nabla_{\|}$ denoting the surface gradient simply reduces to $(\partial/\partial x,\partial/\partial y)$. In this limit, the corresponding energy equation describing heat transfer across the gas/liquid bilayer reduces simply to the 1D Laplace equation $d^2 T/dz^2 = 0$ from which can be derived the temperature distribution along the liquid interface $z=h(\mb{x},t)$:
\begin{equation}
T[h(\mb{x},t)] = T_\mathrm{cold} + \Delta T \, \frac{d_o - h(\mb{x},t)}{d_o + (\kappa - 1)h(\mb{x},t)}.
\end{equation}

The material parameter $\kappa$ denotes the ratio of gas to liquid thermal conductivity evaluated at the temperatures of the respective adjacent substrates. Since the gas layer is always more thermally insulating than the liquid layer, the ratio $\kappa$ is restricted to the range $0 < \kappa <1$. Depending on the materials of choice, however, the magnitude of $\kappa$ can range anywhere from about $1/4$ or higher for molten polymer films like polystyrene overlay by an air film \cite{DT10} to $10^{-4}$ or smaller for liquid metal films such as indium \cite{PW01} overlay by a xenon gas layer \cite{MH16}. The confined geometry leads to self-reinforcing thermocapillary stresses, which promote growth of elongations toward the colder substrate. This process is mitigated only by capillary forces which try to repress formation of regions of high interfacial curvature. Stabilizing gravitational forces, which are orders of magnitude smaller than the thermocapillary forces, are completely negligible. The system described has also been shown to be susceptible to a linear instability \cite{DT09,DT10} which establishes irrespective of the size of the applied thermal gradient. At early times, infinitesimal disturbances generate periodic undulations in film thickness which undergo exponential growth. The fastest growing undulations are characterized by the wavelength
\begin{equation}
\lambda_\mathrm{max}= 2 \pi h_o \left (\frac{4 \gamma_o h_o}{3 \kappa d_o \gamma_T \Delta T} \right)^{1/2}\left(\frac{d_o}{h_o} + \kappa-1 \right)
\end{equation}
where $\gamma_o = \gamma(T_{\mathrm{hot}})$ and $\gamma_T = |d\gamma/dT|_ {T_{\mathrm{hot}}}$. All else equal, a larger difference in temperature $\Delta$ leads to growing undulations of smaller wavelength. Recent \cite{MLT11,FT16,FT18} and ongoing experiments to confirm the mechanism leading to instability so far indicate good agreement with analytic predictions for the fastest growing wavelength and its growth rate.

The dimensionless evolution equation describing the long wavelength thermocapillary model is given by
\begin{equation}
\frac{\partial \widehat{H}}{\partial\hat{\tau}} + \widehat{\nabla}_{\parallel} \cdot
\left \{\frac{\widehat{H}^3}{3 \overline{Ca}} \widehat{\nabla}_{\parallel}\widehat{\nabla}^2_{\parallel} \widehat{H}+
\frac{\kappa \widehat{D} \overline{Ma} \widehat{H}^2}
{2 \big[\widehat{D} +(\kappa-1) \widehat{H} \big]^2}
\widehat{\nabla}_{\parallel} \widehat{H} \right \} = 0
\label{eq:DT10}
\end{equation}
where $\widehat{\mb{X}}=\mb{x}/\lambda_\mathrm{max}$, $\widehat{H}=h(\mb{x},t)/h_o$, $\widehat{D} = d_o/h_o$ and $\hat{\tau} = u_c t/\lambda_\mathrm{max}$, where $u_c$ is chosen to be a characteristic fluid speed based on in-plane thermocapillary flow. Details of the analysis and derivations leading to this form have been previously presented elsewhere \cite{DT10}. The thin film behavior is therefore controlled by two dimensionless numbers, namely a modified Capillary number $\overline{Ca}= \mu u_c/ \epsilon^3 \gamma_o$ and a modified Marangoni number $\overline{Ma}= \epsilon \gamma_T \Delta T/\mu u_c$. These numbers differ from their usual definitions by factors of the small parameter $\epsilon = h_o/\lambda_\mathrm{max}$ intrinsic to the long wavelength approximation. (This parameter should not be confused with the small parameter $\varepsilon$ pertaining to temporal behavior introduced in Section \ref{asymptotics}.)

As evident, Eq. (\ref{eq:DT10}) exhibits a virtual singularity at $H_s =\widehat{D}/(1- \kappa)$ (equivalently in dimensional variables $h=d_o/(1 - \kappa)$). This singularity lies outside the physical domain and beyond the top cold substrate since $\kappa$ is always less than one.  In general, the system described by Eq. (\ref{eq:DT10}) is not limited to initially flat liquid configurations and describes equally well the response  of any initial non-uniform liquid state to thermocapillary forces, in which case $h_o$ denotes the average initial film thickness.

For purposes of this current study, it proves convenient to recast Eq. (\ref{eq:DT10})  into parameter-free form such that
\begin{equation}
\pd[H]{\tau}+ \nbxy \cdot \left[H^3 \, \nabla_{\parallel} \nabla^2_{\parallel}  H+\frac{H^2}{(1-H)^2} \nbxy H\right]=0
\label{eq:mainPDE}
\end{equation}
where $H=\widehat{H}/H_s$, $\textbf{X}=\widehat{\textbf{X}}/X_c$, $\nabla_{\parallel}= X_c \widehat{\nabla}_{\parallel}$ and $\tau=\hat{\tau}/\tau_c$. The scalings for this reduction are given by $X_c = (2 \widehat{D} H_s/3 \kappa \overline{Ma} \, \overline{Ca})^{1/2}$ and $\tau_c=  4 \widehat{D}^2/(3 \kappa^2 H_s \overline{Ma}^2 \,\overline{Ca})$. In this final form, the top cold substrate is located at $H = 1-\kappa$ while the virtual singularity occurs at $H = 1$. Since in this work we wish to investigate the long time behavior of film protrusions which evolve into cuspidal shapes, we restrict attention to small values of $\kappa \simeq 2 \times 10^{-4}$ (characteristic of a liquid metal film overlay by a highly insulating gas layer). This allows for a longer evolution interval not prematurely terminated by contact with the top substrate. Such contact, of course, would modify the fluid behavior in ways not described by Eq. (\ref{eq:mainPDE}).

\section{Stability considerations by analogy to general gradient flows}
\label{gradflowstability}
In previous work \cite{DT09,DT10}, we presented the linear stability analysis of Eq. (\ref{eq:DT10}) which exclusively focused on early time behavior of infinitesimal fluctuations in interfacial temperature or film thickness. That analysis showed that the instability is of Type II \cite{CG09} where all modal fluctuations of wavelength  $\lambda > \lambda_\mathrm{max}/\sqrt{2}$ are linearly unstable irrespective of the value $\Delta T$. Eliciting the stability characteristics of stationary states of the full nonlinear equation given by Eq. (\ref{eq:mainPDE}) requires a different approach based on the system free energy $\mathfrak{F}[H]$. By exploiting an analogy to gradient flows in general, we show next that Eq. (\ref{eq:mainPDE}) does not admit any stable stationary states on a periodic or infinite domain so long as $H > 0$.

\begin{figure}
\centering
\includegraphics[scale=1.2]{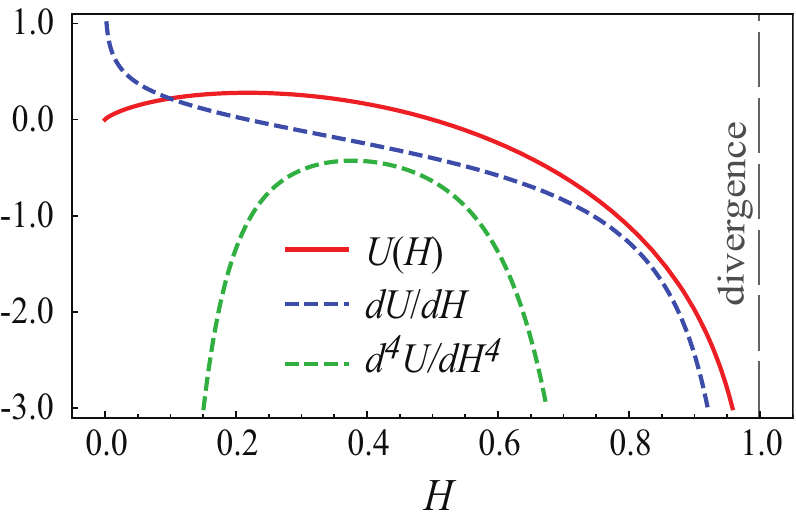}
\caption{Plots of $U(H)$, $0.2\, \times \,\mathrm{d}U/ \mathrm{d}H$ and $0.005\, \times \ \mathrm{d}^4 U/ \mathrm{d}H^4$ for the thermocapillary equation. Magnitudes have been rescaled to accommodate all curves on a common scale.}
\label{fig:uofh}
\end{figure}

Mitlin \cite{M93} has previously shown that the interface equation describing thin film dewetting by van der Waals forces, the process depicted in Figs. \ref{fig:touch_concept} (a) and (b), can be rewritten in Cahn-Hilliard form described by
\begin{equation}
\partial H/\partial \tau = \nbxy\cdot [M(H)\nbxy(\delta \mathfrak{F}/\delta H)]
\label{eq:CHform},
\end{equation}
known more generally as gradient flow form \cite{GO03}. The equation governing the  thermocapillary model can also be written in this form for a free energy functional given by
\begin{equation}
\mathfrak{F}[H,p]=\int_\Omega \, \Big(\frac{1}{2}\left|\nbxy H\right|^2+U(H)\, \Big) \dx{\Omega} - p\, \Big(\int_\Omega H \dx{\Omega}-V\Big),
\label{eq:freeenergywithp}
\end{equation}
with mobility coefficient $M(H)=H^3$, potential function $U(H)=H\ln[(1-H)/H]$ and $\delta \mathfrak{F} / \delta H = -\nbxy^2 H + \mathrm{d}U/\mathrm{d}H$. The curves shown in Fig. \ref{fig:uofh} indicate that $U(H)$ has no global minimum (and that $U(H)$, $\mathrm{d}U/\mathrm{d}H$ and $\mathrm{d}^4 U/\mathrm{d}H^4$ all diverge at the virtual singularity $H = 1$). The energy of the thin film system depends, of course, on the total liquid volume $V$ assumed here to be a conserved quantity. The constraint that the total volume $V = \int_\Omega\ov{H} \dx{\Omega}$ remain constant is enforced through the Lagrange multiplier $p$. As shown in \ref{A1}, $\mathrm{d}\mathfrak{F}/\mathrm{d}\tau\le 0$ on any periodic domain $\Omega$. The proof for an infinite domain simply requires that the integrand in Eq. (\ref{eq:freeenergywithp}) be augmented by a term $U[H(\mb{X} \! \to  \! \infty,\tau)]$, but otherwise proceeds similarly.

We consider stationary solutions $\ov{H}$ represented by the extrema of Eq. (\ref{eq:freeenergywithp}) which satisfy $\delta\mathfrak{F}[\delta H,\delta p; \ov{H},p]=0$ for infinitesimal variations  $\delta H$ and $\delta p$. This yields the value of the Lagrange multiplier
\begin{equation}
p = \left (-\nbxy^2 H + \frac{\mathrm{d}U}{\mathrm{d}H} \right)_{H=\ov{H}} \, ,
\label{eq:equilibrium}
\end{equation}
which reflects the surface pressure required for maintaining stationary states of  constant volume $V$. It has been shown that for a general class of thin film equations \cite{LP02}, which includes the form of Eq. (\ref{eq:mainPDE}), small perturbations to periodic stationary states (i.e. $\delta H \propto \partial^2 \ov{H}/\partial X^2$) lead to negative values of the second variation
\begin{equation}
\delta^2 \mathfrak{F}[\delta H,\delta p; \ov{H},p]=\int_\Omega  |\nbxy \delta H|^2
+\frac{\mathrm{d}^2U}{\mathrm{d}H^2}\Big |_{\ov{H}}\delta H^2  \dx{\Omega}< 0
\label{eq:ddF}
\end{equation}
whenever the potential function satisfies the relation $\mathrm{d}^4 U/\mathrm{d}H^4|_{H \in \ov{H}}<0$. This negative value indicates that there are always nearby states
with same periodicity as $\ov{H}$ of lower free energy. The proof is provided in \ref{A2}. Since for the thermocapillary model the curve $\mathrm{d}^4 U/\mathrm{d}H^4$ shown in Fig. \ref{fig:uofh} is always negative, this therefore proves that Eq. (\ref{eq:mainPDE}) cannot support stable stationary periodic states. This analysis is quite general and can be applied to many other thin film systems (even volume non-conserving systems) so long as the governing interface equation can be cast into the form of Eq. (\ref{eq:CHform}). We next focus on the nonlinear evolution of liquid shapes and show how the thermocapillary model promotes formation of self-similar cusps.

\section{Numerical solution of nonlinear thermocapillary model equation}
\label{numsoln}
To gain insight into the behavior of Eq. (\ref{eq:mainPDE}) in the nonlinear regime, we first examine details of the shapes and dynamics obtained from numerical solutions for both rectilinear [$H(X,\tau)$] and axisymmetric [$H(R,\tau)$] geometry. A mixed Lagrange finite element method \cite{COMSOL} was used to evolve the solutions, subject to no-flux conditions at the origin and end of the computational domain $[0,\lambda_\mathrm{max}/2]$ and initial condition $H(X,\tau=0)= 1/3 \times \left[ 1 + 0.1 \cos (2 \pi X/\lambda_\mathrm{max}) \right]$ (with $X$ replaced by $R$ for the cylindrical case). The restriction to domain size $\lambda_\mathrm{max}/2$ ensured that the dynamics of individual cusp formation could be examined with high resolution without interference from the initial linear instability discussed in Section \ref{tcmodel}. Quadratic elements numbering about 20,000 and of minimum size $4 \times 10^{-8}$ ensured sufficient spatial resolution of the emerging cuspidal region. The mesh sizes were everywhere much smaller than $|\nbxy^2 H|^{-1}$ at all times. Integration in time relied on a second order backward difference scheme with small adaptive time stepping. Typically, full evolution toward the asymptotic shapes required about 11,000 integration steps. Simulations were terminated when the (dimensionless) distance between the virtual singularity at $H=1$ and the liquid cusp apex $H_{\mathrm{apex}}(\tau) = H(0,\tau)$ reached a value of about $10^{-4}$.

\begin{figure}
\centering
\includegraphics[scale = 0.9]{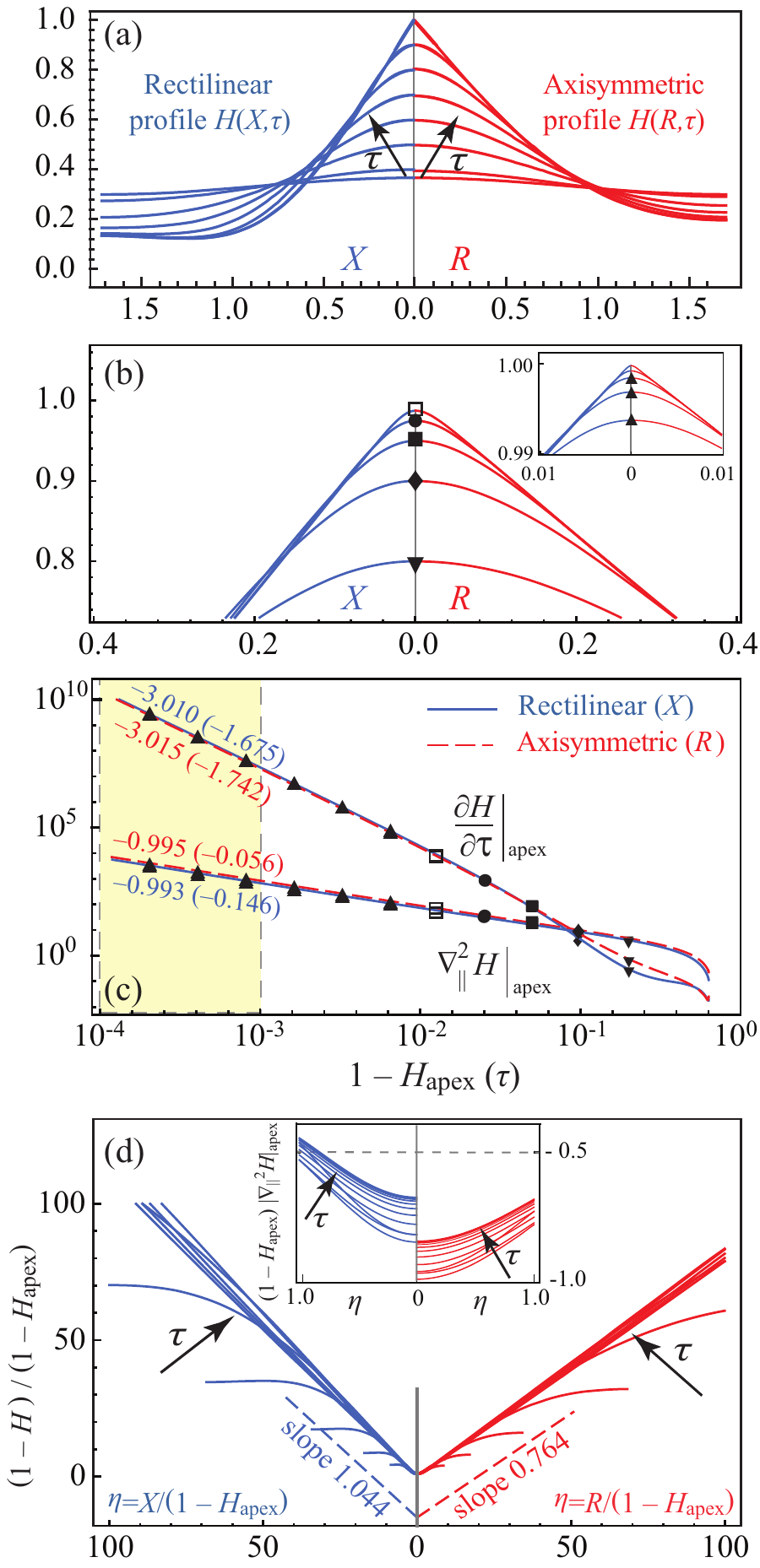}
\caption{Self-similar evolution of conical cusp formation and associated power law exponents for thin film thermocapillary driven system in rectilinear ($X$) and axisymmetric ($R$) geometry. Arrows indicate increasing time $\tau$.
(a) Far field view of cusp formation for $H_{\mathrm{apex}}(\tau) = 0.367, 0.4, 0.5, 0.6, 0.7, 0.8, 0.9, 0.9875$.
(b) Magnified view of conical tip for $H_{\mathrm{apex}}(\tau) = 1 - 0.2/2^n$ showing $n=0$ ($\blacktriangledown$), $n=1$ ($\blacklozenge$), $n=2$ ($\blacksquare$), $n=3$
(\ding{108}) and $n=4$ ($\square$).
Inset: Late time magnified view of conical tip for $H_{\mathrm{apex}}(\tau) = 1 - 0.2/2^n$ showing $n= 5 - 9$ ($\blacktriangle$).
(c) Power law behavior of $\partial H/\partial \tau |_{\mathrm{apex}}$ and
$|\nbxy^2 H|_\mathrm{apex}$ versus $1-H_{\mathrm{apex}}(\tau)$ at the conical apex. Slopes and intercept values (in parentheses) were obtained from least squares fits over the shaded (yellow) region. (d) Rescaled solutions $(1-H)/(1-H_{\mathrm{apex}})$ showing self-similar collapse of the conical tip for $H_{\mathrm{apex}}(\tau) = 1 - 0.2/2^n$ where $n = 0 - 10$. Inset: Rescaled apex curvature $(1-H_{\mathrm{apex}})(\nabla^2_{||} H)_{\mathrm{apex}}$ versus $\eta$ showing self-similar collapse with increasing time.}
\label{fig:sim}
\end{figure}

Shown in Fig. \ref{fig:sim} are far field (a) and magnified views (b) of an evolving cusp capped by a conical tip. As expected from consideration of volume accumulation, the rectilinear geometry leads to a slightly thinner cusp for the same time interval. Inspection of the shape of the fluid tip reveals a conical protrusion with constant slope whose tip radius decreases rapidly in time. Plotted in Fig. \ref{fig:sim} (c) are the tip speed $\partial H/\partial \tau |_\mathrm{apex}$ and magnitude of the tip curvature $|\nbxy^2 H|_\mathrm{apex}$ as a function of the decreasing distance $1-H_{\mathrm{apex}}(\tau)$. The power law behavior observed persists for almost four decades in time indicating robust self-similar growth. The indicated asymptotic values for the slope and intercept values (in parentheses) of the lines shown were obtained from least squares fits over the shaded (yellow) portion shown. This self-similar behavior confirms the relations $\partial H_{\mathrm{apex}}/\partial \tau \sim (1 - H_\mathrm{apex})^{-3}$ and $(\nbxy^2 H)_\mathrm{apex} \sim (1-H_\mathrm{apex})^{-1}$. Introducing the singular time $\tau_s$ where $H_\mathrm{apex} = 1$ - the singular point of Eq. (\ref{eq:mainPDE}) - yields the scaling relations governing the conical tip region, namely $(1-H_\mathrm{apex})/(\tau_s-\tau) \sim (1-H_\mathrm{apex})^{-3}$ and $(1-H_\mathrm{apex})/X^2 \sim (1-H_\mathrm{apex})^{-1}$. These reveal the self-similar variables characterizing this asymptotic regime, namely $X \sim 1-H_\mathrm{apex} \sim (\tau_s-\tau)^{1/4}$, which reflect the lack of an intrinsic spatial or temporal scale  in the conical region. As evident in Fig. \ref{fig:sim}(d), the shape of the conical  tip undergoes collapse onto a common curve when both the vertical and lateral dimensions are normalized by the factor $(1-H_\mathrm{apex})$. The extent of the collapsed region is observed to increase in time. Shown in the inset of Fig. \ref{fig:sim}(d) is the rescaled apical curvature $(1-H_{\mathrm{apex}})(\nabla^2_{||} H)_{\mathrm{apex}}$ versus $\eta = (X,R)/(1-H_{\mathrm{apex}})$, which also exhibits  self-similar collapse. The virtual singularity $H_\mathrm{apex} = 1$ appears therefore to act as an attractor state for formation of the conical tip.

\begin{figure}
\centering
\includegraphics[scale = 1.0]{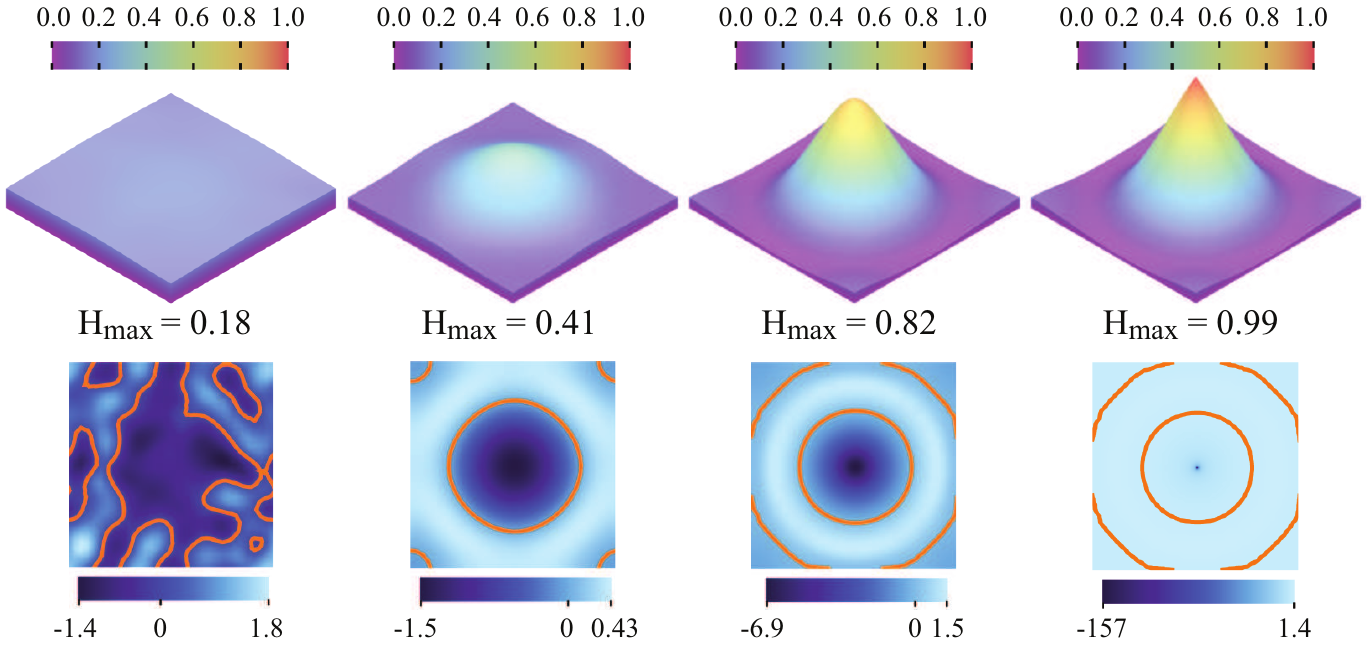}
\caption{Four images of the film thickness $H(\mb{X},\tau)$ (top panel) and interface  curvature $\nabla_\parallel^2 H$ (bottom panel) from numerical simulation of
Eq. (\ref{eq:mainPDE}) on a square periodic domain with edge length
$\lambda_\mathrm{max} \approx 3.02$. The initial condition for this simulation was
$H(\mb{X},0)=\{1-0.05 [\cos(2 \pi X/\lambda_\mathrm{max})+\cos (2 \pi Y/\lambda_\mathrm{max})]+\mathrm{R}(\mb{X})\}/6$, where $\mathrm{R}(\mb{X})$ denotes a  uniformly distributed random variable between -0.2 and 0.2. The maximum film thickness is denoted $H_{\textrm{max}}$. The evolution times depicted are
$\tau = 0.0, 30.0, 50.5 \, \textrm{and} \, 50.84552722$.}
\label{fig:full2d}
\end{figure}

The top panel shown in Fig. \ref{fig:full2d} represents 3D views of an evolving cusp with a conical tip at for the four times designated, as obtained from finite element simulation of the full nonlinear Eq. (\ref{eq:mainPDE}). Additional information about the simulation and accompanying video clip can be found in \ref{videoinfo}. The bottom panel displays the value of the curvature of the gas/liquid interface at every point within the computational domain. The orange curves delineate concave from convex regions. The last image in the bottom panel clearly reveals that the interface evolves into a true cusp capped by a conical tip of decreasing radius flanked by a broader convex surface.

\section{Asymptotic analysis of self-similar cusp formation}
\label{asymptotics}
The exponents extracted from the numerical simulations described in the previous section are also confirmed by analysis of Eq. (\ref{eq:mainPDE}) by considering a Taylor expansion about the virtual singular point $H=1$, which yields the asymptotic evolution equation
\begin{equation}
\pd[H]{\tau} + \nbxy \cdot \left[\nbxy \nbxy^2 H+\frac{1}{(1-H)^2}\nbxy H\right]+\mc{O}(1-H)^{-1}=0~.
\label{eq:leadingPDE}
\end{equation}
Balancing the first and second term with the second and third term yields the same asymptotic relation obtained previously, namely $X \sim 1-H_\mathrm{apex} \sim (\tau_s-\tau)^{1/4}$. Based on these scalings, we introduce the stretched variables
\begin{equation}
\eta=\frac{X}{\varepsilon} \quad \textrm{or} \quad \frac{R}{\varepsilon} \quad \textrm{and} \quad 1 - H = \sum_{n=1}^\infty \varepsilon^n w_n(\eta) \quad \textrm{where} \quad \varepsilon=(\tau_s-\tau)^{1/4}.
\label{eq:similarityvariables}
\end{equation}
Note that if Eq. (\ref{eq:mainPDE}) were truly scale invariant, and not just asymptotically so as $H_\mathrm{apex} \rightarrow 1$, the expansion in Eq. (\ref{eq:similarityvariables}) would terminate at $n=1$. The appearance of the $1-H$ term in the denominator of Eq. (\ref{eq:mainPDE}), however, precludes such global scaling and instead leads to multiscale expansions of the form:
\begin{align}
\pd[H]{\tau}&=\frac{1}{\varepsilon^4}\sum_{n=1}^{\infty}\varepsilon^{n}\mc{T}_n(w_1,\dots,w_n)\label{eq:expandT}\\
\nbxy\cdot \left(H^3\nbxy \nbxy^2 H\right)&=\frac{1}{\varepsilon^4}\sum_{n=1}^{\infty}\varepsilon^{n}\mc{S}_n(w_1,\dots,w_n)\label{eq:expandS}\\
\nbxy\cdot \left[\frac{H^2}{(1-H)^2} \nbxy H\right]&=\frac{1}{\varepsilon^4}\sum_{n=1}^{\infty}\varepsilon^{n}\mc{M}_n(w_1,\dots,w_n)~,\label{eq:expandM}
\end{align}
where the operator symbols $\nbxy$, $\nbxy\cdot$ and $\nbxy^2$ are understood to reduce to the appropriate rectilinear ($X$) or cylindrical ($R$) form for the gradient, divergence and Laplacian operations. To leading order $n=1$, Eq. (\ref{eq:leadingPDE}) then reduces to the nonlinear, fourth order equation given by
\begin{equation}
\mc{T}_1(w_1)+\mc{S}_1(w_1)+\mc{M}_1(w_1)=0~,
\label{eq:leadingODE}
\end{equation}
where the operators $\mc{T}_1$, $\mc{S}_1$ and $\mc{M}_1$ are defined as
\begin{align}
\mc{T}_1(w_1)  &=\frac{1}{4} \left( w_1-\eta \frac{d w_1}{d \eta} \right) \label{eq:T1}\\
\mc{S}_1(w_1)  &=  -\nabla^2_{\eta} \nabla^2_{\eta}w_1 \label{eq:S1}\\
\mc{M}_1(w_1)  &=  \nabla^2_{\eta} \left(\frac{1}{w_1} \right)~.\label{eq:M1}
\end{align}
Here and in what follows, operator subscripts denote differentiation with respect to the self similar variable $\eta$. Required symmetry about the axis of origin yields two boundary conditions, namely $(dw_1/d\eta)_{\eta = 0} = 0$ and $(d^3 w_1/d \eta^3)_{\eta = 0} = 0$. An additional boundary condition is obtained from the requirement that Eq. (\ref{eq:expandT}) remain bounded as $\varepsilon \rightarrow 0$, or equivalently as $\eta \rightarrow \infty$, which requires that the leading term $\mc{T}_1$ vanish. This then leads to the Robin condition $\mc{T}_1(w_1)|_{\eta \to \infty} = 0$. To leading order then, the asymptotic solution to Eq. (\ref{eq:leadingODE}) is satisfied by the Laurent series
\begin{equation}
w^{\infty}_{1}=\sum_{n=1}^{\infty}a_n\eta^{5-4n}=a_1\eta +\mc{O}(\eta ^{-3}) \quad \mathrm{ as }\,\,|\eta|\to \infty~.
\label{eq:asymptote}
\end{equation}

Convergence to $w_1^{\infty}$ can be obtained by linearizing Eq. (\ref{eq:leadingODE}) about the solution $w_1 (\eta \to \infty) = w_1^{\infty}(\eta) + f(\eta)$ which leads to the non-homogeneous linear equation
\begin{equation}
\mc{T}_1(f)+\mc{S}_1(f)-\nabla_\eta^2(f/(w^\infty_1)^2)=0.
\end{equation}
In the limit $|a_1| \ll 1$, this equation leads to a singular perturbation problem whose  inner region is influenced by the fourth order capillary term (not shown). Here we only  focus on the global outer region solutions of the linearized equation obtained by WKBJ analysis where $f(\sigma \eta) = \exp\left[\sigma^{-4/3}\sum_{n=0}^\infty \sigma^{4n/3}S_n(\sigma\eta)\right]$ for $\sigma \ll 1$. Matching terms of order $\sigma^{-4/3}$ and $\sigma^0$, then solving for the resulting two ordinary equations yields the general solution
\begin{equation}
f\sim \beta_0\eta +\sum_{n=1}^3 \frac{\beta_n}{\eta^{\alpha}} \exp
\left[-\frac{3}{4^{4/3}}e^{2 n \pi i/3} \eta^{4/3}\right]+\dots
\label{eq:wkbmodes}
\end{equation}
where $\alpha = 1$ for rectilinear and $\alpha = 5/3$ for axisymmetric geometry. To preclude the first two terms in the summation from undergoing diverging oscillatory behavior, it is required that $\beta_1 = \beta_2 = 0$. The two remaining non-vanishing terms proportional to $\beta_0$ and $\beta_3$ simply reflect an infinitesimal shift in the far field slope and a rapidly decaying function, respectively. Were the analytic solution to Eq. (\ref{eq:leadingODE}) known within the apical region, then the coefficients $\beta_0$ and $\beta_3$ could be obtained by asymptotic matching. Absent that information, the solutions to Eq. (\ref{eq:leadingODE}) are still constrained by the symmetry requirement about at the origin. This constraint imposes that the solutions correspond only to discrete values of the far field slope, as discussed next.

\begin{figure}
\centering
\includegraphics[scale=1.2]{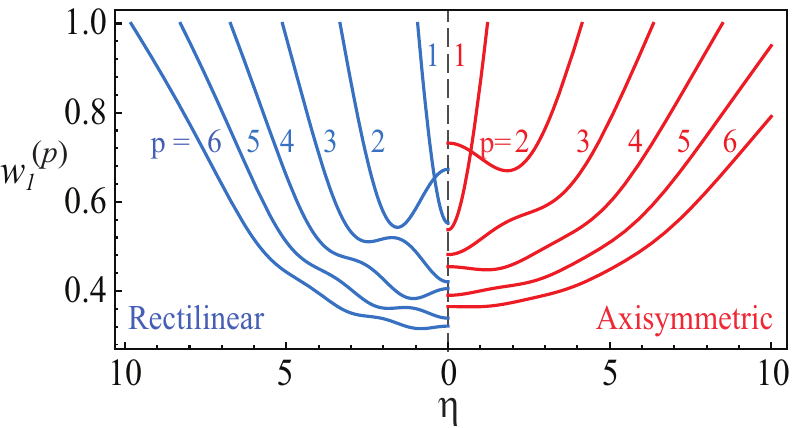}
\caption{Leading order self-similar solutions $w^{(p)}_1$ of Eq. (\ref{eq:leadingODE}). Only the first six convergent solutions are shown.}
\label{fig:planarandaxi}
\end{figure}

The numerical solutions to Eq. (\ref{eq:leadingODE}) were computed on a finite domain  sufficiently long to preclude finite size effects. Simulations with increasing mesh refinement were conducted to assure convergent solutions. Shown in Fig. \ref{fig:planarandaxi} are the first six similarity solutions with selected numerical values listed in Table \ref{tb:firstsix}. The asymptotic interface slopes in the conical  region for axisymmetric geometry are always smaller than the slopes for rectilinear geometry, as expected. The axisymmetric solutions also display weaker oscillatory behavior, likely due to suppression by the capillary pressure associated with the additional term in the interface curvature. The fundamental mode $p=1$ exhibits no oscillatory behavior unlike the higher order solutions $p \geq 2$.

\begin{table}[h]\centering
\caption{\label{tb:firstsix}
Asymptotic values of the interface slope, apex height and apex curvature for the leading order solution $w_1$ to Eq. (\ref{eq:leadingODE}). Numbers in blue and red denote values for rectilinear and axisymmetric geometry, respectively.}
\begin{tabular*}{0.8\textwidth}{@{\extracolsep{\fill}}cccc}
\br
$p$&$\displaystyle{\lim_{\eta \to \infty}} \mathrm{d}w^{(p)}_1/\mathrm{d} \eta$&$ w^{(p)}_1 (0)$ &$\nabla^2_\eta w^{(p)}_1 (0)$\\[-2.5pt]
\mr
1&\bx{1.0437}\quad\rx{0.7639}&\bx{0.5526}\quad\rx{0.5372}&\hphantom{-}\bx{1.2082}\quad\hphantom{-}\rx{1.5563}\\
2&\bx{0.3430}\quad\rx{0.2474}&\bx{0.6728}\quad\rx{0.7317}&\bx{-0.2316}\quad\rx{-0.1624}\\
3&\bx{0.2145}\quad\rx{0.1610}&\bx{0.4204}\quad\rx{0.4816}&\hphantom{-}\bx{0.2021}\quad\rx{-0.1669}\\
4&\bx{0.1580}\quad\rx{0.1196}&\bx{0.4052}\quad\rx{0.4544}&\bx{-0.0884}\quad\rx{-0.0438}\\
5&\bx{0.1257}\quad\rx{0.0962}&\bx{0.3390}\quad\rx{0.3902}&\hphantom{-}\bx{0.0792}\quad\hphantom{-}\rx{0.0526}\\
6&\bx{0.1046}\quad\rx{0.0806}&\bx{0.3211}\quad\rx{0.3649}&\bx{-0.0364}\quad\rx{-0.0087}\\[-2.5pt]\br
\end{tabular*}
\end{table}

Next we compare the fitting coefficients from the asymptotic self-similar analysis of Eq. (\ref{eq:leadingODE}) with those obtained from direct numerical simulations of Eq. (\ref{eq:mainPDE}), which are plotted in Fig. \ref{fig:sim}.
To leading order $1-H_{\mathrm{apex}} \approx \varepsilon w^{(p)}_1(0)$, it can be shown that the intercept value for
$(\partial H / \partial \tau)_{\textrm{apex}}$ is approximately $4\log_{10}[w_1^{(p)} (0)] - \log_{10}4$ and for $(\nbxy^2 H)_{\mathrm{apex}}$ approximately  $\log_{10}[w^{(p)}_1(0)\nbxy^2 w^{(p)}_1(0)]$.
Substitution of the values for $p=1$ from Table \ref{tb:firstsix} into these expressions  yields intercept values for $(\partial H / \partial \tau)_{\textrm{apex}}$
equal to $-1.632$ (rectilinear) and $-1.681$ (axisymmetric). Likewise, the intercept values for $(\nbxy^2 H)_{\mathrm{apex}}$ equal $-0.175$ (rectilinear) and $-0.078$ (axisymmetric). These predicted values are in excellent agreement with the numerical intercept values (shown in parentheses) in Fig. \ref{fig:sim} (c). Additionally, the asymptotic values of the interface slope $\lim_{\eta \to \infty} \mathrm{d}w^{(1)}_1/\mathrm{d} \eta$ given in Table \ref{tb:firstsix} also show excellent agreement when superposed on the profiles in Fig. \ref{fig:sim} (d). The asymptotic values are predicted to be 1.0437 (rectilinear) and 0.7639 (axisymmetric), while the numerical results yield 1.044 and 0.764. Converting back to dimensional form, the value of the interface slope in the region of the conical tip is given by the relation
\begin{equation}
\textrm{Conical tip slope} = (\gamma_T \Delta T/ \gamma_o)^{1/2} [3 \kappa / 2(1 - \kappa)]^{1/2} \times
\lim_{\eta \to \infty} \mathrm{d}w^{(1)}_1/\mathrm{d} \eta ~.
\label{slope}
\end{equation}
With the value of the asymptotic slope known, the remaining parameters in Eq. (\ref{slope}) are set by the material constants of the gas/liquid system of choice and the temperature drop applied to the confining substrates.

In Section \ref{numsoln}, it was shown that the numerical solution to the full nonlinear equation given by Eq. (\ref{eq:mainPDE}) asymptotes to a fluid shape resembling a cusp capped by a conical tip. The asymptotic analysis in this Section reveals that the numerical solution obtained corresponds identically to the fundamental solution $w_1^{(1)}$. A general proof of why the numerical solution always converges to this fundamental solution and not other solutions $w_1^{(p \geq 2)}$ is beyond the scope of this paper. Consideration of this issue by implementing a conventional linear stability analysis of Eq. (\ref{eq:leadingPDE}) is non-trivial due to the multiscale nature of the asymptotic, self-similar base state solutions, which evolve on multiple time scales $\{\varepsilon^n\}^\infty_{n=1}$. Since both the numerical and analytic solutions suggest  that the late stage dynamics of Eq. (\ref{eq:similarityvariables}) is dominated by the term $w^{(p)}_1$, it suffices then to consider infinitesimal perturbations described by
\begin{equation}
1 - H = \varepsilon w^{(p)}_1(\eta) + \varepsilon^{1-4\lambda}\sum_{m=0}^{\infty}e^{im\theta} \phi^{(p)}_m(\eta)\,
\label{axiperturb}
\end{equation}
where $|\phi^{(p)}_m(\eta)| \ll 1$ denotes an infinitesimal modal perturbation to $w^{(p)}_1(\eta)$, $\theta$ is the polar angle in cylindrical coordinates, and $\varepsilon$ is defined in Eq. (\ref{eq:similarityvariables}). The resulting eigenvalue problem is given by
\begin{equation}
\mc{T}_1[\phi^{(p)}_m]+\mc{S}_1 [\phi^{(p)}_m]+\delta \mc{M}_1([\phi^{(p)}_m]=\lambda^{(p)}_m \phi^{(p)}_m,
\label{eq:eigen}
\end{equation}
where $\delta\mc{M}_1[\phi^{(p)}_m]=-\nabla_\eta^2\big[\phi^{(p)}_m/(w^{(p)}_1)^2\big]$ and where differential operators in Eq. (\ref{eq:leadingODE}) have been expanded to include the appropriate $\theta$-dependence. In order for localized perturbations in the far field to preserve constant slope, $\mc{T}_1 [\phi^{(p)}_m]- \lambda^{(p)}_m \phi^{(p)}_m\to 0$ as $\eta \to \infty$. Here, positive eigenvalues $\lambda^{(p)}_m$ reflect perturbations $\phi^{(p)}_m$ with algebraic growth $(\tau_s-\tau)^{1 - 4\lambda}$ faster than the growth $(\tau_s-\tau)^{1/4}$ of the corresponding base state solutions $w_1^{(p)}$. We note that since Eq. (\ref{eq:leadingPDE}) is both space and time translationally invariant, there also exist for each value of $p$ two eigenvalues reflecting these symmetries, namely the eigenfunction $\cos \theta \times d w^{(p)}_1/d\eta$ with eigenvalue 1/4 and the eigenfunction $(w^{(p)}_1 - \eta d w^{(p)}_1)/d\eta)/4$ with eigenvalue 1, respectively.

\begin{figure}
\centering
\includegraphics[scale=1]{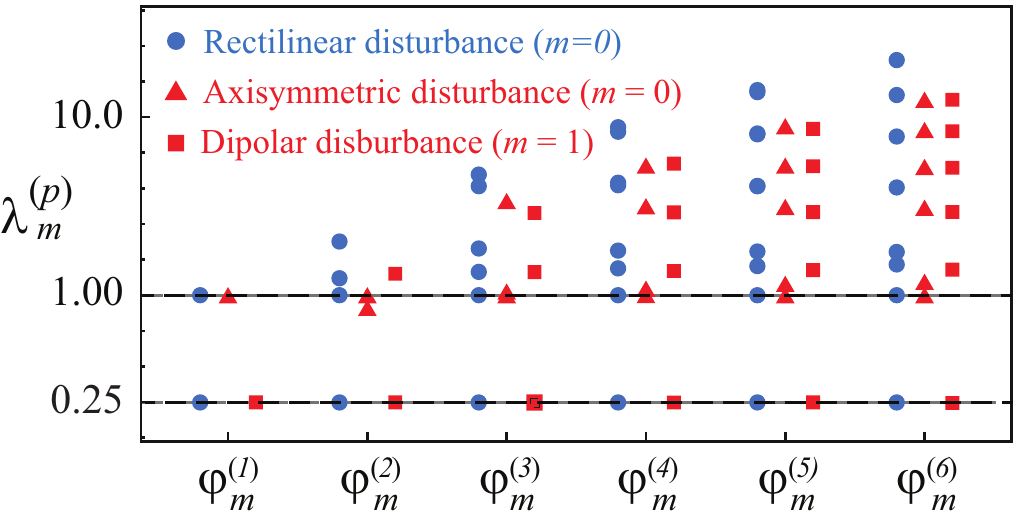}
\caption{Eigenvalue spectrum $\lambda^{(p)}_m$ of Eq. (\ref{eq:eigen}) for perturbations $\phi^{(p)}_m$ to the $w^{(p)}_1$ ($p=1 - 6$) base state solutions of Eq. (\ref{eq:leadingODE}) for rectilinear ($m=0$), axisymmetric ($m=0$) and dipolar ($m=1$) disturbances.}
\label{fig:eigen}
\end{figure}

Plotted in Fig. \ref{fig:eigen} is the eigenvalue spectrum $\lambda^{(p)}_m$ for infinitesimal modal perturbations $\phi^{(p)}_m$ to the first six self-similar base states $w^{(p)}_1$ for $p = 1 - 6$. Each such solution contains $2p$ eigenvalues. Irrespective of the geometry, the fundamental solution $w^{(1)}_1$ is the only solution with no positive eigenvalues aside from 1/4 and 1. The solution $w^{(1)}_1$ is therefore the only solution that is linearly stable to perturbations. The remaining positive eigenvalues increase in magnitude with increasing $p$, indicating more rapid growth and instability associated with the coefficient $\varepsilon^{1-4\lambda}$ multiplying the last term in Eq. (\ref{axiperturb}). The numerical simulations described in Section \ref{numsoln} and plotted in Figs. \ref{fig:sim} and \ref{fig:full2d} were always found to asymptote to the bounded fundamental solution $w^{(1)}_1$. Similar strong convergence to the stable fundamental solution has also been reported for the thin film equation describing van der Waals rupture \cite{WB99} (shown in Fig. \ref{fig:touch_concept}). In that example, initialization of the thin film equation by the corresponding solution $w^{(p \geq 2)}_1$ for that problem leads to a different global liquid film configuration - however, the local behavior in the vicinity of a line or point of rupture nonetheless converges to the fundamental mode $w^{(1)}_1$. A full investigation of the local scaling behavior leading to self-similar cuspidal formation in the thermocapillary system for initial conditions resembling higher order eigenmodes is left for further study. It is anticipated that irrespective of the initial condition, simulation of the full nonlinear evolution equation given by Eq. (\ref{eq:mainPDE}) will still yield film shapes dominated by $w^{(p)}_1$ in the region of the conical tip given that the solution $w^{(1)}_1$ is linearly stable.

\section{Conclusion}
\label{conclusion}
The analysis and simulations presented in this work predict how surface shear forces from self-reinforcing thermocapillary stresses at a gas/liquid interface will produce sharp protrusions resembling cusps capped by conical tip. This finding expands the category of hydrodynamic flows known to form stable cusps to include thin film systems subject to interfacial shear, where the driving force is oriented parallel to the moving interface. The asymptotic analysis reveals how the liquid tip undergoes self-focusing toward a virtual attractor state characterized by a line (rectilinear case) or point (axisymmetric case) singularity via a persistent self-similar process. The asymptotic derivation also yields an simple analytic relation for the slope of the conical tip which should prove useful to experimentalists who may require microarrays with specified tip slope, for beam shaping purposes, design of super antireflective coatings \cite{WA16}, or other applications.

The original system described, based on a thin uniform molten film confined by parallel solid boundaries maintained at different uniform temperature, is known to support a linear instability that forms arrays of rounded protrusions resembling microlenses. These protrusions are expected to evolve into arrays of conical cusps by the nonlinear dynamical process described since the thermal gradient in the region above the fluid tip becomes increasingly large with time, leading to a runaway process. We anticipate that any initial film configuration that contains local maxima in film thickness, whether or not periodically arranged and however initially seeded, will also trigger cusp formation at those locations given the local, self-similar nature of the underlying growth dynamics.

We have previously shown \cite{MLT11,MT11,FT16,FT18} that the evolution process leading to rounded lenslet microarrays can be terminated on demand and the liquid shapes affixed in place by dropping the temperature of both substrates below the solidification point. Rapid solidification of these liquid structures is made possible by two advantageous features: the large surface to volume ratios intrinsic to microscale or nanoscale films which facilitates rapid cooling and digital control over the temperature of the confining substrates. It is fully expected then that similar rapid solidification can be achieved  once the desired conical protrusions have formed in order to solidify and affix their shape on demand. Perhaps alternative methods of flow control by laser manipulation, previously applied to thin film thermocapillary spreading along a solid substrate, can also be used\cite{GGS03}. In summary, the findings and implications outlined in this work offer a novel lithographic method for direct, non-contact fabrication of cuspidal microarrays, whose shapes are more difficult, costly or even impossible to fabricate by other means.

\section*{References}
\normalem
\bibliographystyle{iopart-num}

\providecommand{\newblock}{}

\appendix
\section{Proof of relation $\mathrm{d}\mathfrak{F}[H]/\mathrm{d}\tau \leq 0$}
\label{A1}
We evaluate the quantity $\mathrm{d}\mathfrak{F}[H]/\mathrm{d}\tau$ for the free energy $\mathfrak{F}[H]$ defined in Eq. (\ref{eq:freeenergywithp}) by applying Leibnitz's rule for differentiation over a fixed periodic domain $\Omega$:
\begin{align}
\frac{\mathrm{d}\mathfrak{F}[H]}{\mathrm{d}\tau}
&=\frac{\mathrm{d}}{\mathrm{d}\tau}\int_\Omega \, \Big(\frac{1}{2} \left| \nbxy H \right|^2 + U(H)\Big) \,\mathrm{d} \Omega\\
&=\int_\Omega \, \Big(\nbxy H \cdot \frac{\partial }{\partial\tau}(\nbxy H) + \frac{\mathrm{d}U}{\mathrm{d}H}\frac{\partial H}{\partial \tau} \Big)
\,\mathrm{d} \Omega ~.
\label{pf1:2}
\end{align}
Interchanging the order of operators $\nabla_\parallel$ and $\partial /\partial \tau$ and applying Green's first identity to the first integral in Eq. \ref{pf1:2} gives
\begin{equation}
\frac{\mathrm{d}\mathfrak{F}[H]}{\mathrm{d}\tau}
=\int_\Omega\left(-\nabla_\parallel^2H +\frac{\mathrm{d}U}{\mathrm{d}H}\right)\frac{\partial H}{\partial \tau}
\,\mathrm{d} \Omega~,
\label{integratedform}
\end{equation}
where continuity of $H$ and higher order derivatives ensures that the boundary term proportional to $\nabla_\parallel H$ vanishes identically. Substitution of the term
$\partial H/\partial \tau$ in Eq. (\ref{integratedform}) by the relations given in Eq. (\ref{eq:CHform}) and Eq. (\ref{eq:freeenergywithp}) yields
\begin{equation}
\frac{\mathrm{d}\mathfrak{F}[H]}{\mathrm{d}\tau}
=\int_\Omega\left(-\nabla_\parallel^2H +\frac{\mathrm{d}U}{\mathrm{d}H}\right)
\nabla_\parallel\cdot\left\{ M(H)\nabla_\parallel\left(-\nabla_\parallel^2H +\frac{\mathrm{d}U}{\mathrm{d}H}\right)\right\}
\,\mathrm{d} \Omega ~,
\end{equation}
where $M(H) = H^3$. Application of Green's first identity subject to the vanishing boundary term yields the desired inequality
\begin{equation}
\frac{\mathrm{d}\mathfrak{F}[H]}{\mathrm{d}\tau}
= - \int_\Omega \left\{M(H)\left\lvert\nabla_\parallel\left(-\nabla_\parallel^2H +\frac{\mathrm{d}U}{\mathrm{d}H}\right)\right\rvert^2 \right\}
\,\mathrm{d} \Omega\leq 0~.
\end{equation}

\section{Proof of relation $\delta ^2 \mf{F}[\delta H,\delta p;\ov{H},p] < 0$}
\label{A2}
We consider the free energy associated with a small deviation about a stationary solution $\ov{H}$ of Eq. (\ref{eq:freeenergywithp}) for arbitrary perturbation $\delta H$:
\begin{equation}
\mf{F}[\ov{H}+\delta H] = \mf{F}[\ov{H}]+
\delta\mf{F}[\delta H,\delta p;\ov{H},p]+
\frac{1}{2}\delta ^2 \mf{F}[\delta H,\delta p;\ov{H},p]+\mc{O}(\delta H)^3~.
\end{equation}
By definition, the first variation of the energy $\delta\mf{F}[\delta H,\delta p;\ov{H},p]$ must vanish identically for any such stationary solution $\ov{H}$.
Here, the second variation is given by the integral quantity
\begin{eqnarray}
\delta ^2 \mf{F}[\delta H,\delta p;\ov{H},p]
=\int_\Omega |\nbxy \delta H|^2 +\dd[^2U]{H^2}\Big|_{\ov{H}}\delta H^2- 2\,\delta p \,\delta H\dx{\Omega}
\label{eq:ddf_full}
\end{eqnarray}
subject to the constraint of constant volume $V$ such that
\begin{equation}
\int_\Omega (\ov{H} + \delta H) \dx{\Omega} = V,
\end{equation}
which requires therefore that $\int_\Omega \delta H \dx{\Omega}=0$. This in turn indicates that the integrated value of the last term in Eq. (\ref{eq:ddf_full}) reduces to zero. Application of Green's first identify reduced the second variation $\delta ^2 \mf{F}$ to the form
\begin{equation}
\delta ^2 \mf{F}[\delta H,\delta p;\ov{H},p]
=\int_\Omega \delta H\times\bigg(-\nbxy^2 \delta H +\dd[^2U]{H^2}\Big|_{\ov{H}}\delta H\bigg)\dx{\Omega},
\label{eq:ddf_simp}
\end{equation}
where the additional boundary integral vanishes identically for any periodic perturbation $\delta H$.

It is now a straightforward exercise to show that there always exist admissible arbitrary perturbations $\delta H$ such that $\delta ^2 \mf{F}[\delta H,\delta p;\ov{H},p]$ is always strictly negative. We recall from Eq. (\ref{eq:equilibrium}) that the interfacial pressure $p$ (i.e. Lagrange multiplier) corresponding to a stationary state $\ov{H}$ of volume V is given by
\begin{equation}
p = \left (-\nbxy^2 H + \frac{\mathrm{d}U}{\mathrm{d}H} \right)_{H=\ov{H}}
\label{eq:ss}
\end{equation}
Differentiating Eq. (\ref{eq:ss}) twice with respect to $X$ yields the relation
\begin{equation}\label{eq:ssid}
-\nbxy^2 \pd[^2\ov{H}]{X^2}+ \dd[^2U]{H^2}\Big|_{\ov{H}}\pd[^2\ov{H}]{X^2}=- \dd[^3U]{H^3}\Big|_{\ov{H}}\Big(\pd[\ov{H}]{X}\Big)^2.
\end{equation}
Substituting Eq. (\ref{eq:ssid}) into Eq. (\ref{eq:ddf_simp}) for perturbations of the form $\delta H = \partial^2 \ov{H}/\partial X^2$ with vanishing total volume yields
\begin{eqnarray}
\delta ^2 \mf{F}[\delta H,\delta p;\ov{H},p]
&=-\int_\Omega\Big(\pd[^2\ov{H}]{X^2}\Big)\dd[^3U]{H^3}\Big|_{\ov{H}}\Big(\pd[\ov{H}]{X}\Big)^2 \dx{\Omega}\\
&=-\int_\Omega\dd[^3U]{H^3}\Big|_{\ov{H}}\frac{1}{3}\pd[]{X}\Big(\pd[\ov{H}]{X}\Big)^3 \dx{\Omega}\\
&=\frac{1}{3}\int_\Omega\Big(\dd[^4U]{H^4}\Big|_{\ov{H}}\pd[\ov{H}]{X}\Big)\Big(\pd[\ov{H}]{X}\Big)^3 \dx{\Omega}\\
&=\frac{1}{3}\int_\Omega\Big(\pd[\ov{H}]{X}\Big)^4 \dd[^4U]{H^4}\Big|_{\ov{H}}\dx{\Omega}.
\label{last}
\end{eqnarray}
All boundary terms from integrations by parts vanish due to periodic boundary conditions. For the thermocapillary model described by Eq. (\ref{eq:freeenergywithp}), the potential function $U(H)=H\ln[(1-H)/H]$ for $H \in (0,1)$ and therefore
\begin{equation}
\dd[^4U]{H^4}= - \ \frac{2 (1-2H)^2 + 4H^2}{H^3 (1-H)^4} < 0.
\end{equation}
When substituted into Eq. (\ref{last}), this yields the relation $\delta ^2 \mf{F}[\delta H,\delta p;\ov{H},p] < 0$. This inequality assures that for every nonuniform stationary state $\ov{H}$ such that $\partial H/\partial X$ is not everywhere zero, there always exists a neighboring state $\ov{H}+\delta H$ of lower free energy.

\section{Description of Video Clip}
\label{videoinfo}

The video clip shows images obtained by finite element simulation of the parameter-free interface equation [Eq. (4) in text] given by
\begin{equation}
\label{eq:pde}
\pd[H]{\tau}+\nabla\cdot\left[H^3\nabla^3H+\frac{H^2}{(1-H)^2}\nabla H\right]=0~.
\end{equation}
The initial condition was chosen to be
\begin{equation}
H(\mb{X},0)=\{1-0.05 [\cos(2 \pi X/\lambda_\mathrm{max})+\cos (2 \pi Y/\lambda_\mathrm{max})]+\mathrm{R}(\mb{X})\}/6~,
\end{equation}
where $\mathrm{R}(\mb{X})$ denotes a uniformly distributed random variable between -0.2 and 0.2. The designations $H_{\textrm{max}}$ and $H_{\textrm{min}}$ in each image denote  the maximum and minimum value in film thickness within the computational domain at the designated time. The simulation was performed on a square periodic domain of edge length $\lambda_\mathrm{max} = 2\pi / K_\textrm{max} \approx 3.02$ where
\begin{equation}
K_\textrm{max}=\frac{1}{(1-H_\mathrm{avg})\sqrt{2H_\mathrm{avg}} }
\end{equation}
and $H_\mathrm{avg}$ is the initial average film thickness, here chosen to be  $H_\mathrm{avg}=1/6$ to yield $\lambda_\textrm{max}\sim 3.02$.

\begin{figure}
\centering
\includegraphics[scale = 0.5]{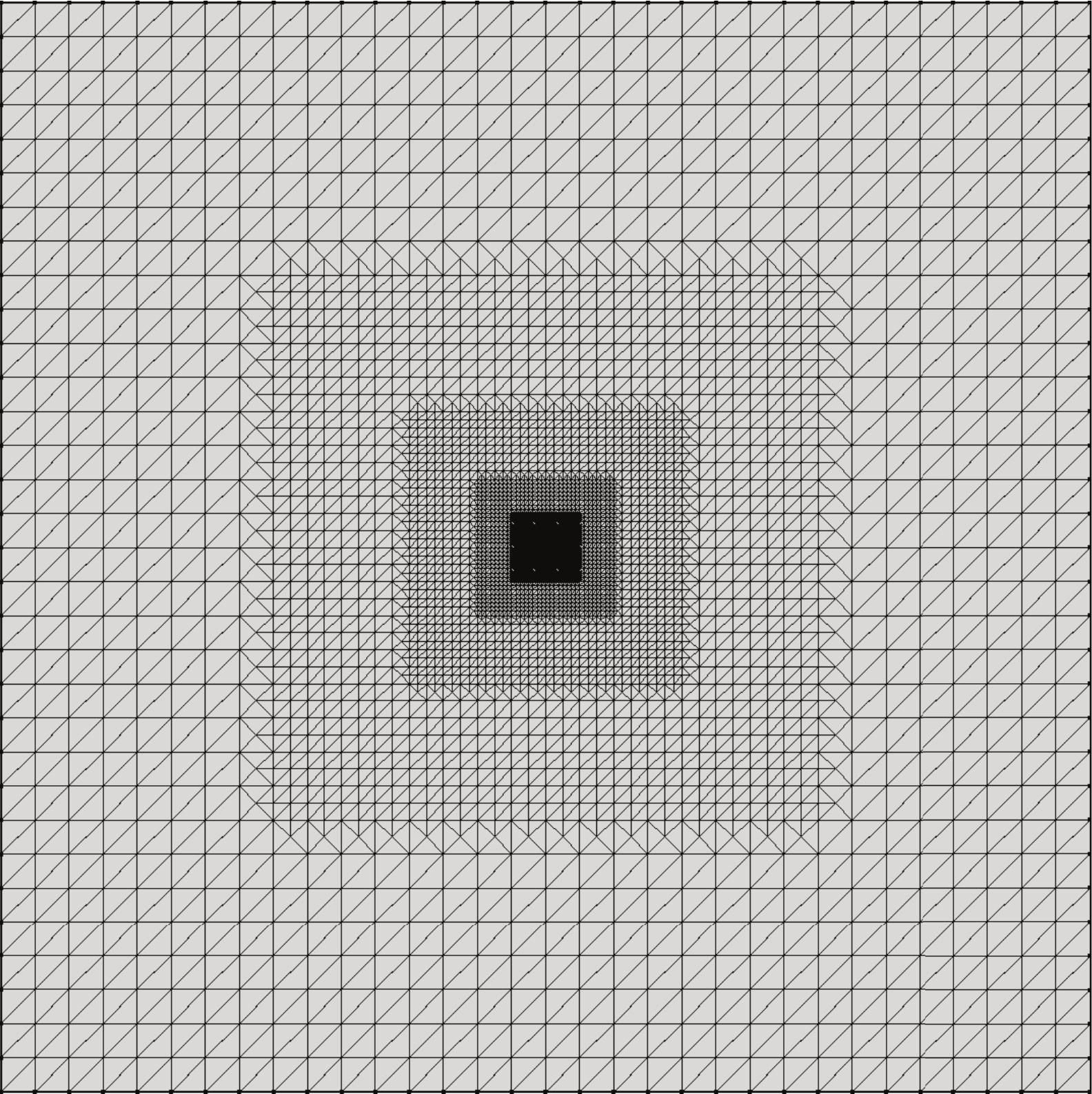}
\caption{Image of progressively refined mesh used to resolve details in the apical region.}
\label{fig:mesh}
\end{figure}

The square domain was discretized into 15,872 triangular elements of quadratic order with 63,746 degrees of freedom in total. Since the evolving cusp was centered about the origin of the domain, the nested mesh shown in Fig. \ref{fig:mesh} was implemented in order to resolve details of the apical region with sufficient resolution.
The edge size of the smallest mesh element in the central was about 0.0004,
intentionally chosen to be smaller than the minimum value of $|\nabla_\parallel H|^{-1}\sim \mathcal{O}(10^{-2})$ throughout the simulation time.

The finite element simulation was run until the cusp apex height reached a maximum value $H_\textrm{max}$ that just exceeded 0.99. To better resolve the progressively faster growth, the video clip spans three distinct time intervals, concatenated back to back, namely
\begin{align*}
\text{Stage I: }& 00.000 \leq \tau \leq 50.000 \quad \textrm{with} \quad \Delta \tau = 1.00\times 10^{-1}\\
\text{Stage II: }& 50.005 \leq \tau \leq 50.845 \quad \textrm{with} \quad\Delta \tau = 5.00\times 10^{-2}\\
\text{Stage III: }& 50.845 < \tau \leq 50.845522 \quad \textrm{with} \quad \Delta \tau = 5.22\times 10^{-6}. &
\end{align*}

\end{document}